# Na$_4$IrO$_4$: Singular manifestation of square-planar coordination of a transition metal in $d^5$ configuration due to weak on-site Coulomb interaction

Sudipta Kanungo[b], Binghai Yan[b,c], Patrick Merz[b], Claudia Felser,[b] and Martin Jansen*[a,b]

*In memoriam Rudolf Hoppe.*

**The square-planar coordination of transition metals has been assumed to require the $d^8$ or $d^9$ electron configuration so far. Here we report a square-planar structure of the IrO$_4$ entity with a $d^5$ electron configuration for Na$_4$IrO$_4$. The weak Coulomb interaction of Ir-5$d$ states is found to stabilize this unconventional square-planar structure.**

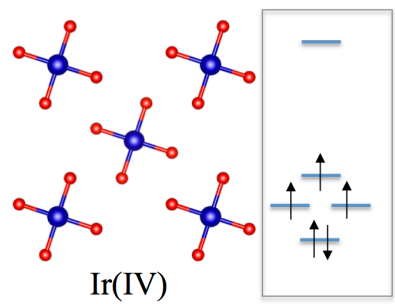

**Abstract:** Local environments and valence electron counts primarily determine the electronic states and physical properties of transition metal complexes. For example, square-planar surroundings found in transition oxometalates such as cuprates are usually associated with the $d^8$ or $d^9$ electron configuration. In this work, we address an exotic square-planar mono-oxoanion [IrO$_4$]$^{4-}$ as observed in Na$_4$IrO$_4$ with Ir(IV) in $d^5$ configuration, and characterize the chemical bonding by experiment and *ab initio* calculations. We find that Na$_4$IrO$_4$ in its ground state evolves a square-planar coordination for Ir(IV) because of the weak Coulomb repulsion of Ir-5$d$ electrons. In contrast, in its 3$d$ counterpart, Na$_4$CoO$_4$, Co(IV) is in tetrahedral coordination, due to strong electron correlation. Na$_4$IrO$_4$ thus may serve as a simple paradigmatic platform for studying the ramifications of Hubbard type Coulomb interactions on local geometries.

Tetrahedral geometry is by far the prevailing coordination geometry encountered with isolated entities [MO$_4$]$^{n-}$ and is electrostatically favored. Consequently, other coordination geometries such as square planar require special local electronic configurations and covalent bonding contributions to be stabilized. It has been quite easy to rationalize the occurrence of square-planar coordination, even in purely qualitative terms, from electron counting. The $d^8$ and $d^9$ electron configurations are frequently associated to quadrate surroundings of transition metals,[1] while for main group elements, the combination of four covalently bonded ligands and the presence of two lone pairs stringently directs toward such a topology (VSEPR model).[2] Recent work on oxygen-depleted perovskites has somewhat blurred this clear picture. Using soft chemistry routes, a square-planar local geometry has been realized for Fe(II), for example, in SrFeO$_2$.[3] However, this compound is reported to be metastable, and its structure does not represent the ground-state configuration. Moreover, the coordination polyhedra are not solitary, but linked via vertices to form 2D sheets. It is thus difficult to judge whether the local geometry is intrinsically stable or rather, is supported by extended lattice effects.

The family of $A_4$IrO$_4$ iridates(IV), synthesized previously with $A$ = Na, K and Cs [4] by Rudolf Hoppe, and featuring the first examples of square-planar *mono*-oxoanions for a transition metal with an electron configuration different from $d^8$ or $d^9$, is raising deeper concern in this context. As a particularly puzzling fact, Co$^{4+}$ in the lighter homologue, Na$_4$CoO$_4$, is as expected tetrahedrally coordinated in the high-spin $d^5$ state.[5] Therefore, it is compelling to investigate why Ir$^{4+}$ can be stabilized in the square-planar structure, rather than tetrahedral. Furthermore it would be interesting to determine the factors that induce different local structures for IrO$_4$ and CoO$_4$ entities in the same oxidation state, and whether they can be attributed as an effect of spin-orbit coupling (SOC), Coulomb interactions or others. Moreover, how these different structures affect the characteristic electronic and magnetic properties.

To unravel the apparent conundrum, we revisit this exotic class of square-planar iridates (IV) by validating earlier structural work, followed by an in depth theoretical analysis of the chemical bonding. Our current study is further motivated by the fact that oxoiridates have attracted much attention in recent

[a] Prof.Dr. Martin Jansen
Max-Planck-Institutfür Festkorperforschung
D-70569 Stuttgart, Germany
E-mail: m.janse@fkf.mpg.de
[b] Dr.SudiptaKanungo, Dr. Binghai Yan, Patrick Merz, Prof. Dr. Claudia Felser, Prof.Dr. Martin Jansen
Max-Planck-Institut für Chemische Physik fester Stoffe
D-01187 Dresden, Germany
[c] Dr. Binghai Yan
Max-Planck-Institut für Physik komplexer Systeme
D-01187 Dresden, Germany

Supporting information for this article is given via a link at the end of the document.

years, owing to the interesting emergent phenomena induced by the cooperative effect of Coulomb interactions and SOC, such as spin liquids,[6] topological Mott insulators[7] and Weyl semimetals[8].

All $A_4IrO_4$ compounds with A = Na, K and Cs[4] share the same $IrO_4$ square-planar local environment that determines the electronic and magnetic properties. The Na iridate crystallizes in a tetragonal lattice whereas the K and Cs compounds display lower symmetric monoclinic lattices. Without loss of generality, we chose high symmetric $Na_4IrO_4$ and its 3d counterpart $Na_4CoO_4$ as model systems in this work.

We have prepared $Na_4IrO_4$ as a pure, single-phase powder, applying the azide/nitrate route.[9] Using Rietveld profile fitting,[10] the earlier crystal structure determination has been corroborated, confirming the presence of the ideally square-planar $IrO_4$ entity. $Na_4IrO_4$ crystallizes in the tetragonal space group I4/m,[4] (Figure 1), where the perfect $IrO_4$ square-planar environment exists in the ab plane. In contrast, $Na_4CoO_4$ exhibits a triclinic lattice where the $[CoO_4]^{4-}$ oxoanion is tetrahedral,[5] as shown in Figure 1c.

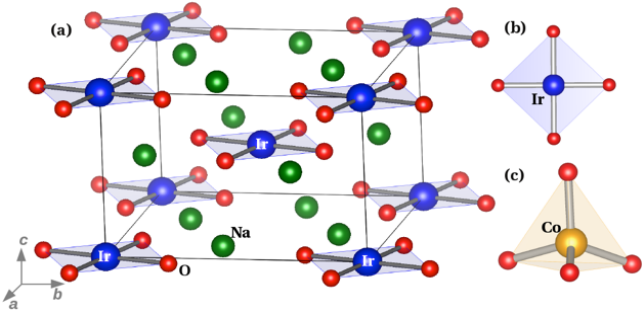

**Figure 1.** (a) Crystal structure of $Na_4IrO_4$. Green, blue and red spheres represent Na, Ir and O atoms, respectively. (b) The square-planar structure of the $IrO_4$ entity, where the Ir-O bond length is $d_{Ir-O}$ = 1.902 Å. (c) The tetrahedral environment of the $CoO_4$ entity (Co represented by orange sphere) in $Na_4CoO_4$, where the average Co-O bond length is $\bar{d}_{Co-O}$ = 1.732 Å.

The temperature dependence of the magnetic susceptibility ($\chi$) exhibits clear antiferromagnetic (AFM) ordering for $Na_4IrO_4$ at low temperature. As shown in Figure 2, the AFM transition occurred at the Neel temperature, $T_N$ ~ 25 K. The inverse of the susceptibility ($\chi^{-1}$) deviated slightly from linearity in the low temperature region. This is plausibly related to the population of multiplet states, which is beyond the scope of the current work. Fitting the inverse susceptibility in the high temperature region (100 < T < 400 K) according to the Curie-Weiss law, given by $\chi$ = $C/(T-\theta)$ where C and $\theta$ are the Curie and Weiss constants, respectively, we obtained $\theta$ = −74 K, C = 1.47 emu Kmol$^{-1}$, and $\mu_{eff}$ = $2.83\sqrt{C}$ = 3.43 $\mu_B$. The determined effective magnetic moment is similar to the ideal moment of the S = 3/2 spin-state $\mu_{eff} = 2\sqrt{S(S+1)} = 3.87 \mu_B$, as confirmed in the following theoretical ab initio investigation.

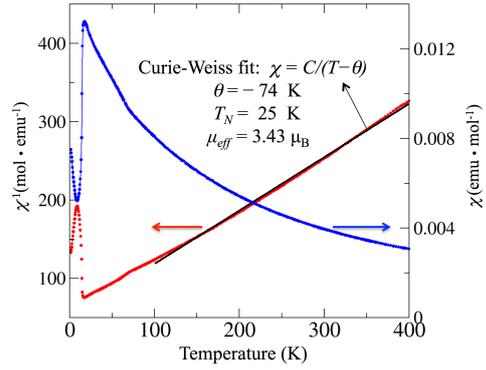

**Figure 2.** Temperature dependence of the magnetic susceptibility measured at 3.5 T field for $Na_4IrO_4$. Red and blue dots stand for $\chi^{-1}$ and $\chi$ respectively. The black line is the fitting of $\chi^{-1}$ to the temperature according to the Curie-Weiss law in the temperature range of 100-400 K.

The electronic and magnetic properties were further investigated using density-functional theory (DFT) calculations within the framework of generalized gradient approximation (GGA).[11] The correlation effect was included in the GGA+U [12] scheme for transition metal elements. In the GGA calculations, we first fully relaxed the lattice parameters and atomic positions of $Na_4IrO_4$ and $Na_4CoO_4$. The resultant optimized structures correlated well with experimental values (Table S2 in SI [10]). When we included the Coulomb on-site interaction U to Ir-5d and Co-3d states, the ground state structures remained unchanged for both compounds. For example, the square-planar configuration remained as the ground state of $Na_4IrO_4$ even up to an unphysically large value, $U_{Ir}$ = 6 eV. In addition, a tetrahedral $[IrO_4]^{4-}$ oxoanion exists in a metastable phase only at a very large $U_{Ir}$ value (e.g. 6 eV). If we reduced $U_{Ir}$ to values smaller than 4 eV, this tetrahedral oxoanion collapsed quickly to the square-planar analogue in the structure optimization. Such a lattice instability indicated the significance of effective Coulomb interactions in determining the structure of $Na_4IrO_4$.

The d orbital splitting under the crystal field (CF) and corresponding electron occupation were extracted from calculations, as shown in Figure 3. In the square-planar environment of the $IrO_4$ entity, Ir ($d^5$) exhibits an S = 3/2 spin state, while in the tetrahedral environment of the $CoO_4$ entity, Co ($d^5$) displays an S = 5/2 spin state. Both cases opened moderate energy gaps at the Fermi energy. However, when trapping Ir ($d^5$) in the metastable structure in tetrahedral CF, a filling of $e^4 t_2^1$ occurred due to the weak Coulomb U of Ir-5d states. Consequently, considerable density of states (DOS) appeared at the Fermi energy (Figure 4b), leading to a Jahn-Teller type instability. Therefore, differences in structural coordination of $[IrO_4]^{4-}$ and $[CoO_4]^{4-}$ oxoanions can be rationalized by differences in the on-site Coulomb interaction of Ir-5d and Co-3d states.

As shown in Figure 4a, an energy gap of 0.5 eV between spin-up $d_{xy}$ and spin-down $d_{xz,yz}$ was observed for $Na_4IrO_4$.

From the DOS, we were able to estimate the CF splitting of the square-planar environment of Ir-5$d$ states (Figure 3), which accommodates five $d$ electrons. According to Hund's rule, the first four $d$ electrons fill the lowest four levels in the spin-up channel. However, the fifth electron cannot fill the $d_{x^2-y^2}$ level (not shown in Figure 4a) owing to the large CF energy gap (~ 4 eV) between $d_{xy}$ and $d_{x^2-y^2}$. It occupies the $d_{3z^2-r^2}$ state in the spin-down channel at the expense of the electron pairing energy of 1 eV, which is estimated from the energy split between spin-down and spin-up $d_{3z^2-r^2}$ levels. Thus, a final $S = 3/2$ state is realized, explaining the experimentally observed magnetic moment. In the above calculations, we presumed the ferromagnetic (FM) order in the lattice. We also performed calculations with the AFM spin configuration, in which spins at the body center and corner sites were oppositely orientated. The AFM phase was found to be 25 meV per formula unit lower in energy than the FM phase. However, AFM and FM configurations exhibited the same pattern of CF splitting and orbital filling without any qualitative differences. For the sake of simplicity, we used the FM configuration for all further calculations presented in this work. In addition, the main effect of SOC was found to lift the degeneracy of $d_{xz,yz}$ bands, but it did not affect the main features of the orbital diagram, as well as the structure instability. Therefore, we omitted SOC in the following discussions (see more details on SOC in SI [10]).

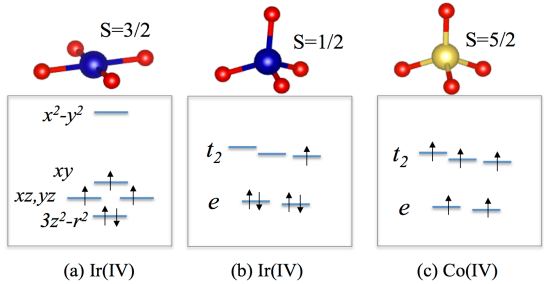

**Figure 3.** Crystal field splitting of $d$ orbitals and corresponding occupation for (a) the square-planar IrO$_4$ group, (b) tetrahedral IrO$_4$ group in the artificially fixed tetrahdral Na$_4$IrO$_4$ structure and (c) tetrahedral CoO$_4$ group.

When $U_{Ir}$ < 4 eV was applied in GGA+U calculations, which is the reasonable small $U$ region for Ir-5$d$ states, the square-planar structure remains as the ground state phase, as discussed above. Surprisingly, when a large $U_{Ir}$ (e.g. 6 eV) is employed, a *metastable* structure with a tetrahedral IrO$_4$ entity was really obtained. Such tetrahedral environment exhibits $e-t_2$ type CF levels, which are filled as $e^2 t_2^3$ due to the artificially large $U_{Ir}$. However, if we apply $U_{Ir}$ < 4 eV to this metastable tetrahedral structure, the same CF levels can only be filled as $e^4 t_2^1$ (Figure 3b). Then these partially filled $t_2$ bands induce considerable DOS at the Fermi energy (Figure 4b), although the degeneracy of $t_2$ states split slightly due to four non-equivalent Ir-O bonds in the tetrahedral structure. Consequently, a Jahn-Teller type instability emerges and causes the IrO$_4$ entity to become square-planar, resulting in an energy gap. Therefore, the relative weak $U_{Ir}$ was the principal origin of the instability in the tetrahedral structure of the [IrO$_4$]$^{4-}$ oxoanion.

For Na$_4$CoO$_4$, the CF levels are occupied as $e^2 t_2^3$ with the $S = 5/2$ high-spin state, owing to the large Coulomb $U$ of Co-3$d$ states. In the GGA calculation, this already leads to an energy gap (Figure 4c). At the large value of $U_{Co}$ (e.g. 5 eV) in GGA+U calculations, which is known to be a good approximation for Co, above scenario still holds. In addition, we note that the square-planar structure is a *metastable* phase for Na$_4$CoO$_4$, which was found to be energetically 0.4 eV per formula unit higher than the tetrahedral structure in the GGA calculations. Following the trend from 3$d$ to 5$d$ systems, we would expect that the corresponding 4$d$ material Na$_4$RhO$_4$ possibly present an intermediate state between Na$_4$CoO$_4$ and Na$_4$IrO$_4$, although Na$_4$RhO$_4$ has yet to be synthesized in experiment. GGA calculations reveal that Na$_4$RhO$_4$ exhibits distorted tetrahedral environment of the RhO4 group with an $S = 1/2$ spin state. Even though partially filled $t_2$ states appear at the Fermi energy, which is similar to Na$_4$IrO$_4$, a slight distortion of the RhO4 tetrahedra is enough to open an energy gap of 0.1 eV (GGA). This effect is caused by the narrow bandwidth of Rh-$t_2$ states, while the bandwidth of Ir-$t_2$ states is too wide to open an energy gap. So Na$_4$RhO$_4$ can be stabilized at the $S = 1/2$ state while Na$_4$IrO$_4$ cannot. In all, Ir, Rh and Co compounds constitute an interesting phase diagram, wherein the electron correlation effect interplays with the crystal structure and magnetic properties. It will be noteworthy that materials at critical points between different phases may be synthesized from solid solutions of two compounds *e.g.* as Na$_4$Rh$_x$Ir$_{1-x}$O$_4$, which probably promise exotic phenomena such as magnetoelastic coupling and electron-lattice coupling.

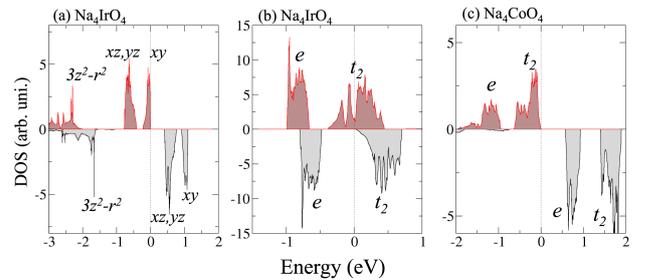

**Figure 4.** Density of states for (a) square-planar Na$_4$IrO$_4$, (b) tetrahedral Na$_4$IrO$_4$ and (c) tetrahedral Na$_4$CoO$_4$ calculated within GGA. The Fermi energy is shifted to zero. Positive (red curves) and negative (black curves) values represent spin-up and spin-down channels, respectively.

In summary, we experimentally confirm solitary, ideally square planar mono-oxoanions [IrO$_4$]$^{4-}$ to be constitutive components of the ground state structure of Na$_4$IrO$_4$. We have rationalized this unprecedented local coordination of a transition metal with a $d^5$ electron configuration by quantum chemical calculations on the GGA + U level, which have revealed the low interatomic Coulomb repulsion to be the pivotal factor of influence stabilizing the square planar coordination of a central cation in $5d^5$ electron configuration. Because of the weak $U$ of Ir-$5d$ electrons, the square-planar coordination geometry is the favorable one for $[IrO_4]^{4-}$, while in contrast, due to strong $U$ of Co-$3d$ states, the $[CoO_4]^{4-}$ oxoanion, e.g. in $Na_4CoO_4$, adopts the tetrahedral geometry associated with a high spin $S = 5/2$ state. These findings are introducing the strength of Coulomb repulsion, commonly expressed as Hubbard $U$, as another independent pattern of explanation for electronically driven distortions of local topologies, in addition to the Jahn-Teller effect and spin-orbit coupling.[15]

## Experimental Section

For the preparation of $Na_4IrO_4$, the following starting materials were used: sodium azide (Alfa Aesar, 99%), sodium nitrate (Merck, 99.99%) and iridium oxide $IrO_2$, which was obtained after heating of $IrCl_3*xH_2O$ (Alfa Aesar, 99.9%) at 650°C for 24 h, under oxygen flow. The starting materials were mixed according to the following equation, using 2.5% excess of the sodium compounds, $10NaN_3 + 2NaNO_3 + 3IrO_2 \rightarrow 3Na_4IrO_4 + 16N_2$ in an agate mortar into a fine powder and pressed into pellets (ø 13mm, 5*10$^4$ N,5min). After drying overnight under vacuum (10$^{-3}$ mbar) at 120°C, the pellets were placed under argon in a special steel vessel provided with a silver inlay. [9] Under slow argon flow, the following temperature treatment was applied: 25 °C→ 250 °C (100 °C/h), 250 °C → 380 °C (5 °C/h), 380 °C → 600 °C (20 °C/h) and subsequent annealing for 50 h at 600°C before cooling down to room temperature at a rate of −20 °C/h. The as-obtained black microcrystalline powder appears red after further cleaving under dried kerosene, and decomposes in water forming a dark blue solution. Since the samples are sensitive to humid air, they were sealed in glass ampoules under argon and all manipulations were performed in a strict inert atmosphere of purified argon. Laboratory X-ray powder diffraction (XRPD) studies were performed with X'Pert PRO diffractometer (PANalytical B.V., Netherlands) diffractometer Cu-$K_{\alpha1}$ radiation (λ= 0.154056 nm). XPRD was recorded at room temperature in 2θ range 10 to 120°. The magnetization was measured in the temperature range from 0.8 to 400 K using a Quantum design MPMS-XL7 SQUID magnetometer.

## Theoretical calculations

DFT calculations were performed within the plane-wave basis set based on pseudo-potentials as implemented in the Vienna *ab-initio* simulation package (VASP).[14] The GGA exchange-correlation functional was employed following the Perdew-Burke-Ernzerhof prescription.[11] The on-site electron-electron repulsion beyond the GGA was taken into account through GGA+U [12] calculations. For the plane-wave basis, a 500 eV plane-wave cutoff was applied. A k-point mesh of 8 × 8 × 6 in the Brillouin zone was used for self-consistent calculations. In the structural optimization, the atomic positions and the lattice parameters, both were fully optimized without constrains. Positions of the atoms were relaxed toward equilibrium until the force became less than 0.01 eV/Å. Hybrid-functional calculations were also employed to validate the GGA+U results.[10]

## Acknowledgements

Acknowledgement is made to the ERC Advanced Grant (291472) and the Deutsche Forschungsgemeinschaft (DFG) through SFB 1143. B.Y. thanks the helpful discussion with P. Adler.

**Keywords:** Electronic structure • Transition metals • Coordination geometry • Ab initio calculations • Oxometalates

# Supplementary Information

# Na$_4$IrO$_4$: Singular manifestation of square-planar coordination of a transition metal in *d$^5$* configuration due to weak on-site Coulomb interaction

Sudipta Kanungo[b], Binghai Yan[b,c], Patrick Merz[b], Claudia Felser,[b] and Martin Jansen*[a,b]

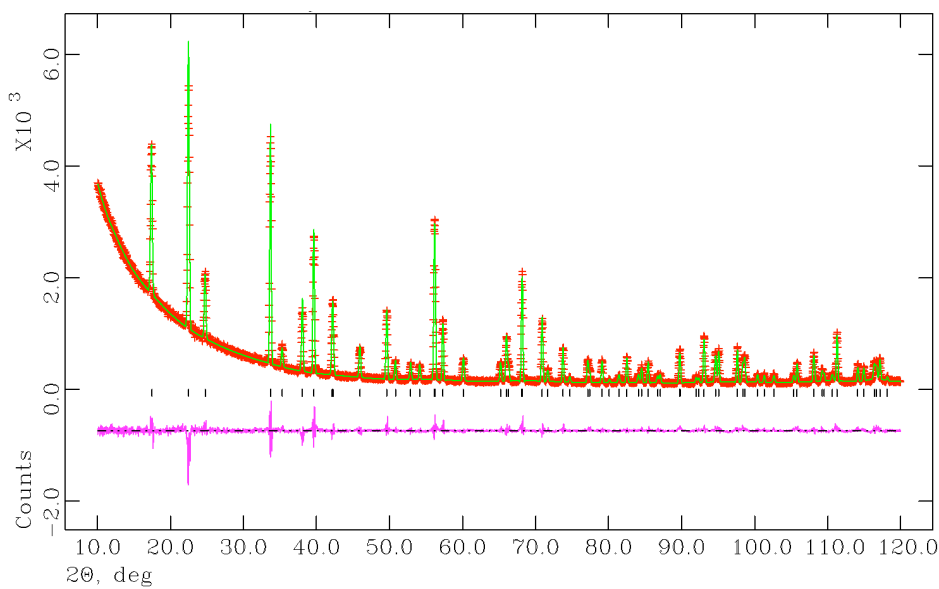

**Fig. S1:** X ray powder diffraction pattern of Na$_4$IrO$_4$ (red points: observed; green line: fit from Rietveld refinement in I4/m; pink line: difference curve; lower black bars, Bragg peaks.

**Table S1**: Results of the crystal structure refinements of Na$_4$IrO$_4$ as obtained from laboratory x-ray. In the tetragonal structure with I4/m the atoms are in the following positions: Na in 8c (x,y,0), Ir in 2a (0,0,0), O in 8h (x,y,0) respectively.

| Crystal Symmetry | Space Group | a (Å) | c (Å) | V (Å$^3$) | R$_{wp}$ | R$_p$ | $\chi^2$ |
|---|---|---|---|---|---|---|---|
| Tetragonal | I4/m | 7.1840(4) | 4.7249(5) | 243.85(1) | 6.6 % | 4.3 % | 2.343 % |

[a]  Prof.Dr. Martin Jansen
     Max-Planck-Institutfür Festkorperforschung
     D-70569 Stuttgart, Germany
     E-mail: m.janse@fkf.mpg.de
[b]  Dr.SudiptaKanungo, Dr. Binghai Yan, Patrick Merz, Prof. Dr. Claudia Felser, Prof.Dr. Martin Jansen
     Max-Planck-Institut für Chemische Physik fester Stoffe
     D-01187 Dresden, Germany
[c]  Dr. Binghai Yan
     Max-Planck-Institut für Physik komplexer Systeme
     D-01187 Dresden, Germany

**Table S2** Crystal structure for three compounds $Na_4IrO_4$, $Na_4CoO_4$ and $Na_4RhO_4$. For $Na_4IrO_4$ and $Na_4CoO_4$ structures are compared between experimental and theoretically optimized structures. Last column of the table contains the theoretically optimized crystal structure for the $Na_4RhO_4$ compounds, which is yet to be synthesized. Both GGA and hybrid functional (HSE06) calculations reach the same conclusions, where the crystal symmetries and the spin state remain the same between GGA and HSE06 calculations (also see Fig. S4). The only difference between the GGA and HSE06 is that the volume and the subsequent bond lengths are smaller in the case of HSE06 than that of GGA. We note that GGA results are closer to the experimental values.

|  | $Na_4IrO_4$ (Sp. Gr. = I4/m) | | | $Na_4CoO_4$ (Sp. Gr.=P2) | | | $Na_4RhO_4$ (Sp. Gr.=P1) | |
|---|---|---|---|---|---|---|---|---|
|  | Experiment [a] | Theoretical optimized | | Experiment [b] | Theoretical optimized | | Theoretical optimized | |
|  |  | GGA | Hybrid functional [c] |  | GGA | Hybrid functional | GGA | Hybrid functional |
| Lattice parameters (Å) | a=b=7.184 c= 4.7249 | a=b=7.207 c=4.704 | a=b=7.076 c=4.624 | a=5.700, b=5.723, c=8.640 | a=5.692, b=5.704, c=8.665 | a=5.584, b=5.600, c=8.504 | a= 6.697, b= 6.017, c=8.498 | a= 6.547, b= 5.911, c=8.400 |
| Volume (Å$^3$) | 243.85 | 244.33 | 231.53 | 247.86 | 247.74 | 234.22 | 250.33 | 238.20 |
| (Ir/Co/Rh)-O bond lengths (Å) | Ir-O=1.942 | Ir-O=1.938 | Ir-O=1.912 | Co-O =1.811, 1.763, 1.856, 1.822 | Co-O= 1.838, 1.807, 1.845, 1.837 | Co-O= 1.799 1.772 1.805 1.799 | Rh-O=1.885, 1.935, 1.966, 1.967 | Rh-O=1.865, 1.938, 1.926, 1.908 |

[a] our synthesized crystal structure as mentioned in table S1.
[b] M. Jansen, Z. Anorg. Allg. Chem. **1975**, 417, 35; C. Jeannot, B. Malaman, R. Gerardin, B. Oulladiaf, J. Solid State Chem. **2002**, 165, 266.
[c] J. Heyd, G. E. Scuseria, M. Ernzerhof, J. Chem. Phys. **2003**, 118, 8207; J. Heyd, G. E. Scuseria, M. Ernzerhof, J. Chem. Phys. **2006**, 124, 219906.

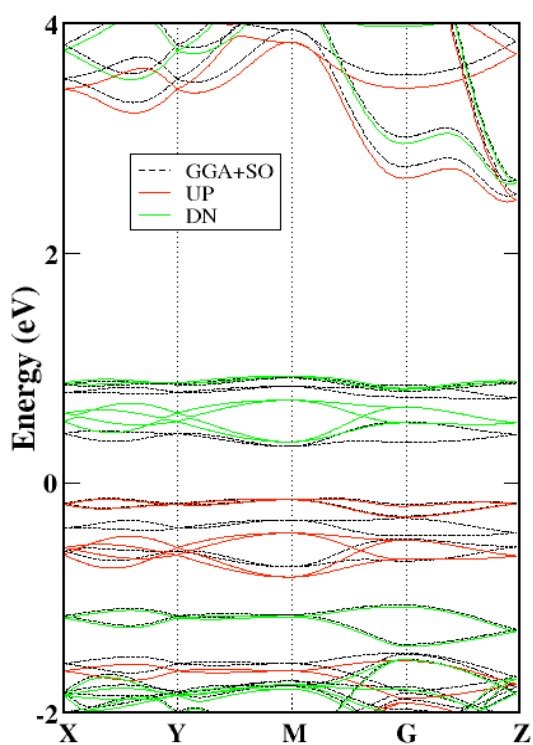

**Fig.S2:** GGA band structure of $Na_4IrO_4$ w/o [red (dark solid) and green (light solid)] and with spin-orbit

coupling [black (dark dash)] for $Na_4IrO_4$ compound. Plot shows the main features of the band structures near the Fermi level remains unaltered with and w/o the spin-orbit coupling.

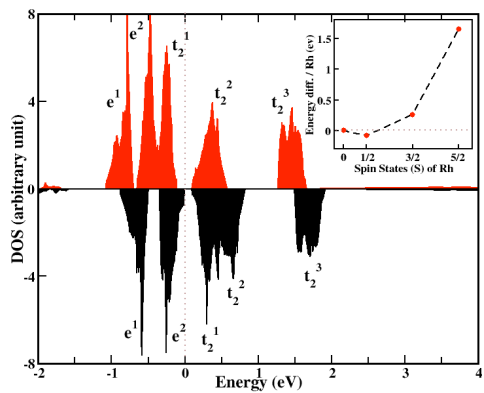

**Fig. S3:** DOS of $Na_4RhO_4$ from spin-polarized GGA calculations, projected onto Rh-4$d$ states. DOS shows $e$-$t_2$ splitting under the tetrahedral environment, with completely filled $e$ in both majority and minority spin channels and partially filled $t_2$ in the majority spin channel (red curves) and completely empty $t_2$ in minority spin channel (black curves). DOS along with the magnetic moment (Rh=0.54 $\mu_B$ / site), suggests that Rh (IV) - $4d^5$ are in the S = ½ state (LS). Inset shows total energy difference for different spin states at the Rh site. Plot shows that S=½ spin state is the lowest energy state compared to the other possible spin states. We checked the robustness of this S = ½ state in the tetrahedral environment increasing on-site Hubbard U. We found that up to $U_{eff}$ = 3 eV, S = ½ state is the ground state for the $Na_4RhO_4$. This S = ½ state in the tetrahedral environment, being stabilized due to small bandwidth of Rh-4d bands near the Fermi levels and opens up a very tiny gap ~ 0.1 eV.

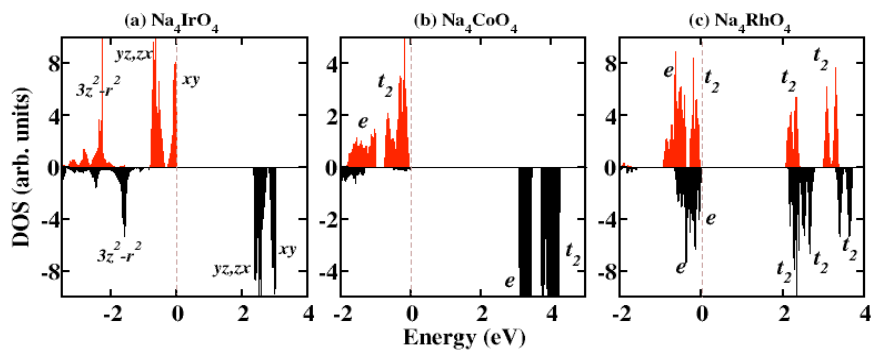

**Fig. S4:** DOS for the three compounds (a) $Na_4IrO_4$, (b) $Na_4CoO_4$, (c) $Na_4RhO_4$ from spin polarized hybrid functional (HSE06) calculations. DOS is projected onto Ir-5$d$, Co-3$d$ and Rh-4$d$ states respectively, using the structures mentioned in Table S2. Interestingly the major features in the electronic structures, i.e. the crystal symmetry, splitting of $d$ states and the spin state remain unchanged between GGA and HSE06 results for the all three compounds. It should be noted that the gap at the Fermi level is significantly enhanced in HSE06 compared to GGA results, which is expected due to the strong correlation taken into HSE06. HSE06 calculations validate our main conclusions based on GGA and GGA+U.